\documentclass[twocolumn]{aastex61}
\pdfoutput=1 
\usepackage{amsmath,amstext}
\usepackage[T1]{fontenc}
\usepackage{apjfonts}
\usepackage{amsmath}
\usepackage{ulem}
\usepackage[figure,figure*]{hypcap}
\usepackage{csvsimple}

\shorttitle{HARPS-N Solar RVs Are Dominated By Large, Bright Magnetic Regions}
\shortauthors{T. Milbourne et al.}

\begin{document}


\title{HARPS-N Solar Radial-Velocity Variations Are Dominated By Large, Bright Magnetic Regions}

\author{T. W. Milbourne}
\affiliation{Department of Physics, Harvard University, 17 Oxford Street, Cambridge MA 02138, USA}
\affiliation{Harvard-Smithsonian Center for Astrophysics, Cambridge, MA 02138, USA}
\author{R. D. Haywood}
\altaffiliation{NASA Sagan Fellow}
\affiliation{Harvard-Smithsonian Center for Astrophysics, Cambridge, MA 02138, USA}
\author{D. F. Phillips}
\affiliation{Harvard-Smithsonian Center for Astrophysics, Cambridge, MA 02138, USA}
\author{S. H. Saar}
\affiliation{Harvard-Smithsonian Center for Astrophysics, Cambridge, MA 02138, USA}
\author{H. M. Cegla}
\altaffiliation{CHEOPS Fellow, SNSF NCCR-PlanetS}
\affiliation{Observatoire de Gen\`eve, Universit\'e de Gen\`eve, 51 chemin des Maillettes, 1290 Versoix, Switzerland}
\author{A. C. Cameron}
\affiliation{Centre for Exoplanet Science, SUPA, School of Physics and Astronomy, University of St Andrews, St Andrews KY16 9SS, UK}
\author{J. Costes}
\affiliation{Astrophysics Research Centre, School of Mathematics and Physics, Queen`s University Belfast, Belfast, BT7 1NN, UK}
\author{X. Dumusque}
\affiliation{Observatoire de Gen{\'e}ve, 51 Ch. des Maillettes, 1290 Sauverny, Switzerland}
\author{N. Langellier}
\affiliation{Department of Physics, Harvard University, 17 Oxford Street, Cambridge MA 02138, USA}
\affiliation{Harvard-Smithsonian Center for Astrophysics, Cambridge, MA 02138, USA}
\author{D. W. Latham}
\affiliation{Harvard-Smithsonian Center for Astrophysics, Cambridge, MA 02138, USA}
\author{J. Maldonado}
\affiliation{INAF-Osservatorio Astronomico di Palermo, Piazza del Parlamento 1, 90134 Palermo, Italy}
\author{L. Malavolta}
\affiliation{INAF-Osservatorio Astronomico di Padova, Vicolo dell`Osservatorio 5, 35122 Padova, Italy}
\affiliation{Dipartimento di Fisica e Astronomia ``Galileo Galilei'', Universit{\`a} di Padova, Vicolo dell`Osservatorio 3, I-35122 Padova, Italy}
\author{A. Mortier}
\affiliation{Astrophysics Group, Cavendish Laboratory, J.J. Thomson Avenue, Cambridge CB3 0HE, UK}
\affiliation{Centre for Exoplanet Science, SUPA, School of Physics and Astronomy, University of St Andrews, St Andrews KY16 9SS, UK}
\author{M. L. Palumbo III}
\affiliation{Harvard-Smithsonian Center for Astrophysics, Cambridge, MA 02138, USA}
\affiliation{Department of Astronomy \& Astrophysics, The Pennsylvania State University, University Park, PA 16802, USA}
\author{S. Thompson}
\affiliation{Astrophysics Group, Cavendish Laboratory, J.J. Thomson Avenue, Cambridge CB3 0HE, UK}
\author{C. A. Watson}
\affiliation{Astrophysics Research Centre, School of Mathematics and Physics, Queens University Belfast, Belfast BT7 1NN, UK}
\author{F. Bouchy}
\affiliation{Observatoire de Gen\`eve, Universit\'e de Gen\`eve, 51 chemin des Maillettes, 1290 Versoix, Switzerland}
\author{N. Buchschacher}
\affiliation{Observatoire de Gen\`eve, Universit\'e de Gen\`eve, 51 chemin des Maillettes, 1290 Versoix, Switzerland}
\author{M. Cecconi}
\affiliation{INAF-Fundacion Galileo Galilei, Rambla Jose Ana Fernandez Perez 7, E-38712 Brena Baja, Spain}
\author{D. Charbonneau}
\affiliation{Harvard-Smithsonian Center for Astrophysics, Cambridge, MA 02138, USA}
\author{R. Cosentino}
\affiliation{INAF-Fundacion Galileo Galilei, Rambla Jose Ana Fernandez Perez 7, E-38712 Brena Baja, Spain}
\author{A. Ghedina}
\affiliation{INAF-Fundacion Galileo Galilei, Rambla Jose Ana Fernandez Perez 7, E-38712 Brena Baja, Spain}
\author{A. G. Glenday}
\affiliation{Harvard-Smithsonian Center for Astrophysics, Cambridge, MA 02138, USA}
\author{M. Gonzalez}
\affiliation{INAF-Fundacion Galileo Galilei, Rambla Jose Ana Fernandez Perez 7, E-38712 Brena Baja, Spain}
\author{C-H. Li}
\affiliation{Harvard-Smithsonian Center for Astrophysics, Cambridge, MA 02138, USA}
\author{M. Lodi}
\affiliation{INAF-Fundacion Galileo Galilei, Rambla Jose Ana Fernandez Perez 7, E-38712 Brena Baja, Spain}
\author{M. L\'opez-Morales}
\affiliation{Harvard-Smithsonian Center for Astrophysics, Cambridge, MA 02138, USA}
\author{C. Lovis}
\affiliation{Observatoire de Gen{\'e}ve, 51 Ch. des Maillettes, 1290 Sauverny, Switzerland}
\author{M. Mayor}
\affiliation{Observatoire de Gen{\'e}ve, 51 Ch. des Maillettes, 1290 Sauverny, Switzerland}
\author{G. Micela}
\affiliation{INAF-Osservatorio Astronomico di Palermo, Piazza del Parlamento 1, 90134 Palermo, Italy}
\author{E. Molinari}
\affiliation{INAF-Fundacion Galileo Galilei, Rambla Jose Ana Fernandez Perez 7, E-38712 Brena Baja, Spain}
\affiliation{ INAF-Osservatorio Astronomico di Cagliari, Via della Scienza 5-09047 Selargius CA, Italy}
\author{F. Pepe}
\affiliation{Observatoire de Gen{\'e}ve, 51 Ch. des Maillettes, 1290 Sauverny, Switzerland}
\author{G. Piotto}
\affiliation{INAF-Osservatorio Astronomico di Padova, Vicolo dell`Osservatorio 5, 35122 Padova, Italy}
\affiliation{Dipartimento di Fisica e Astronomia ``Galileo Galilei'', Universit{\`a} di Padova, Vicolo dell`Osservatorio 3, I-35122 Padova, Italy}
\author{K. Rice}
\affiliation{SUPA, Institute for Astronomy, Royal Observatory, University of Edinburgh, Blackford Hill, Edinburgh EH93HJ, UK}
\affiliation{Centre for Exoplanet Science, University of Edinburgh, Edinburgh, UK}
\author{D. Sasselov}
\affiliation{Harvard-Smithsonian Center for Astrophysics, Cambridge, MA 02138, USA}
\author{D. S{\'e}gransan}
\affiliation{Observatoire de Gen{\'e}ve, 51 Ch. des Maillettes, 1290 Sauverny, Switzerland}
\author{A. Sozzetti}
\affiliation{INAF-Osservatorio Astrofisico di Torino, via Osservatorio 20, 10025 Pino Torinese, Italy}
\author{A. Szentgyorgyi}
\affiliation{Harvard-Smithsonian Center for Astrophysics, Cambridge, MA 02138, USA}
\author{S. Udry}
\affiliation{Observatoire de Gen{\'e}ve, 51 Ch. des Maillettes, 1290 Sauverny, Switzerland}
\author{R. L. Walsworth}
\affiliation{Department of Physics, Harvard University, 17 Oxford Street, Cambridge MA 02138, USA}
\affiliation{Harvard-Smithsonian Center for Astrophysics, Cambridge, MA 02138, USA}

\begin{abstract}
State of the art radial-velocity (RV) exoplanet searches are currently limited by RV signals arising from stellar magnetic activity. 
We analyze solar observations acquired over a 3-year period during the decline of Carrington Cycle 24 to test models of RV variation of Sun-like stars. A purpose-built solar telescope at the High Accuracy Radial velocity Planet Searcher for the Northern hemisphere (HARPS-N) provides disk-integrated solar spectra, from which we extract RVs and $\log{R'_{\rm HK}}$. The Solar Dynamics Observatory (SDO) provides disk-resolved images of magnetic activity. The Solar Radiation and Climate Experiment (SORCE) provides near-continuous solar photometry, analogous to a \emph{Kepler} light curve. 
We verify that the SORCE photometry and HARPS-N $\log{R'_{\rm HK}}$ correlate strongly with the SDO-derived magnetic filling factor, while the HARPS-N RV variations do not. To explain this discrepancy, we test existing models of RV variations. We estimate the contributions of the suppression of convective blueshift and the rotational imbalance due to brightness inhomogeneities to the observed HARPS-N RVs. We investigate the time variation of these contributions over several rotation periods, and how these contributions depend on the area of active regions. We find that magnetic active regions smaller than $60 \ \rm Mm^2$ do not significantly suppress convective blueshift. Our area-dependent model reduces the amplitude of activity-induced RV variations by a factor of two. The present study highlights the need to identify a proxy that correlates specifically with large, bright magnetic regions on the surfaces of exoplanet-hosting stars.
\end{abstract}


\section{Introduction}
The radial velocity (RV) method is a powerful tool for exoplanet detection and mass estimation (\citealt{FirstDetection_1995}, \citealt{Fischer2016}). When used in conjunction with transit measurements, determined using observations from CoRoT, \emph{Kepler}, \emph{K2}, and TESS (\citealt{CoRoT_2009}, \citealt{Borucki977}, \citealt{1538-3873-126-938-398}, \citealt{TESS2014}), the RV method allows for determinations of planetary densities, relevant to studies of the internal structure of detected planets \citep{ZengSasselov2013}. The reflex RV amplitude induced by an Earth-mass planet in the habitable zone of a Sun-like star is about 10 cm~s$^{-1}$, the target sensitivity of next-generation spectrographs \citep{Pepe_et_al_2010}. However, RV measurements are currently dominated by the effects of stellar activity. In particular, acoustic oscillations, granulation  and supergranulation due to surface magneto-convection, and other stellar activity processes contribute to RV signals exceeding 1 m~s$^{-1}$ as discussed by \cite{Saar_et_al_1997}, \cite{schrijver_zwaan_2000}, \cite{Meunier_et_al_2010}, \cite{Dumusque2011}, and \cite{Meunier_et_al_2015}. These activity processes must be understood to successfully interpret current observations.

Stellar activity processes act on distinct timescales. Over periods of a few minutes, stellar p-modes (that is, the propagation of acoustic vibrations) are dominant \citep{Broomhall_et_al_2009}. The upward and downward motion of convecting plasma also contributes to the overall RV signal; these granulation processes (including supergranulation and mesogranulation) are dominant over periods between hours and a few days (\citealt{Palle_et_al_1995}, \citealt{DelMoro_2004}, \citealt{Kjeldsen_et_al_2005}). Contributions from magnetic features, such as dark sunspots, bright photospheric plage (i.e., the magnetically laced photosphere under chromospheric plage) and photospheric network, dominate on timescales longer than a rotation period (\citealt{Meunier_et_al_2010}, \citealt{Haywood_et_al_2016}). Since exoplanet surveys often target low-activity stars, the behavior of stars near activity minimum must be considered to ensure accurate RV detections of low-mass exoplanets.
 
The close proximity of the Sun makes it an ideal test case for studying different stellar signals and correlating with RV measurements. Numerous ground and space-based instruments, such as Global Oscillations Network Group (GONG, \citealt{HARVEY1988117}), the SOlar Radiation and Climate Experiment (SORCE, \citealt{Rottman2005}), and the Solar Dynamics Observatory (SDO, \citealt{Pesnell2012}), perform detailed observations of the Sun's surface. In parallel with these instruments, a custom-built solar telescope installed at the Telescopio Nazionale Galileo (TNG) on La Palma makes disk-integrated, spectroscopic measurements of the Sun as a star using the state-of-the-art High Radial velocity Planet  Searcher for the Northern Hemisphere (HARPS-N) spectrograph (\citealt{Xavier_2015}, \citealt{Phillips2016}). This allows us to observe the Sun as we would any other star in high-precision RV exoplanet surveys. By comparing this rich data set with solar photometry and disk-resolved images, we investigate the contributions of different activity processes to RV measurements, and how these contributions vary over different timescales.


Previously, \cite{Meunier_et_al_2010_MDI} reconstructed the solar RV using images from the Michelson Doppler Imager (MDI) onboard the Solar and Heliospheric Observatory (SOHO). They also investigated how the convective RV shift scales with active region area. In another work, the authors investigate the relative contributions of large versus small active regions to the solar RV \citep{Meunier_et_al_2010}.  \cite{Haywood_et_al_2016} used data from the Helioseismic and Magnetic Imager (HMI, \citealt{Schou2012}) onboard SDO to reconstruct the magnetically-driven solar RVs, and compared these to HARPS RVs derived from sunlight reflected off the asteroid Vesta. Both these works, however, suffer from practical limitations: \cite{Meunier_et_al_2010_MDI} were limited by the spatial resolution of MDI, and were unable to measure the impact of small active regions on the activity-driven RV. While \cite{Haywood_et_al_2016} used higher-resolution HMI images in their work, their observations of Vesta only spanned 70 days, approximately 2.5 synodic solar rotation periods. In order to fully characterize the effects of magnetic activity on the solar RVs, we need to understand the contributions of large and small active regions, and how these contributions evolve over the course of many rotation periods.

In this work, we use contemporaneous disk-averaged solar telescope spectra, HMI solar images, and SORCE Total Irradiance Monitor (TIM) measurements of the Total Solar Irradiance (TSI) (\citealt{Kopp2005a}, \citealt{Kopp2005b}) taken between July 2015 and September 2017, near the end of Solar Cycle 24, to estimate how the RV contributions from convective and photometric solar magnetic activity vary over several solar rotation periods, approaching solar minimum (late 2018/early 2019). We also investigate how these contributions vary with the size of the active regions producing these RV perturbations.

\section{Observations}
\label{sec:Obs}
\begin{figure*} 
\begin{center} 
\includegraphics[width=.95\textwidth]{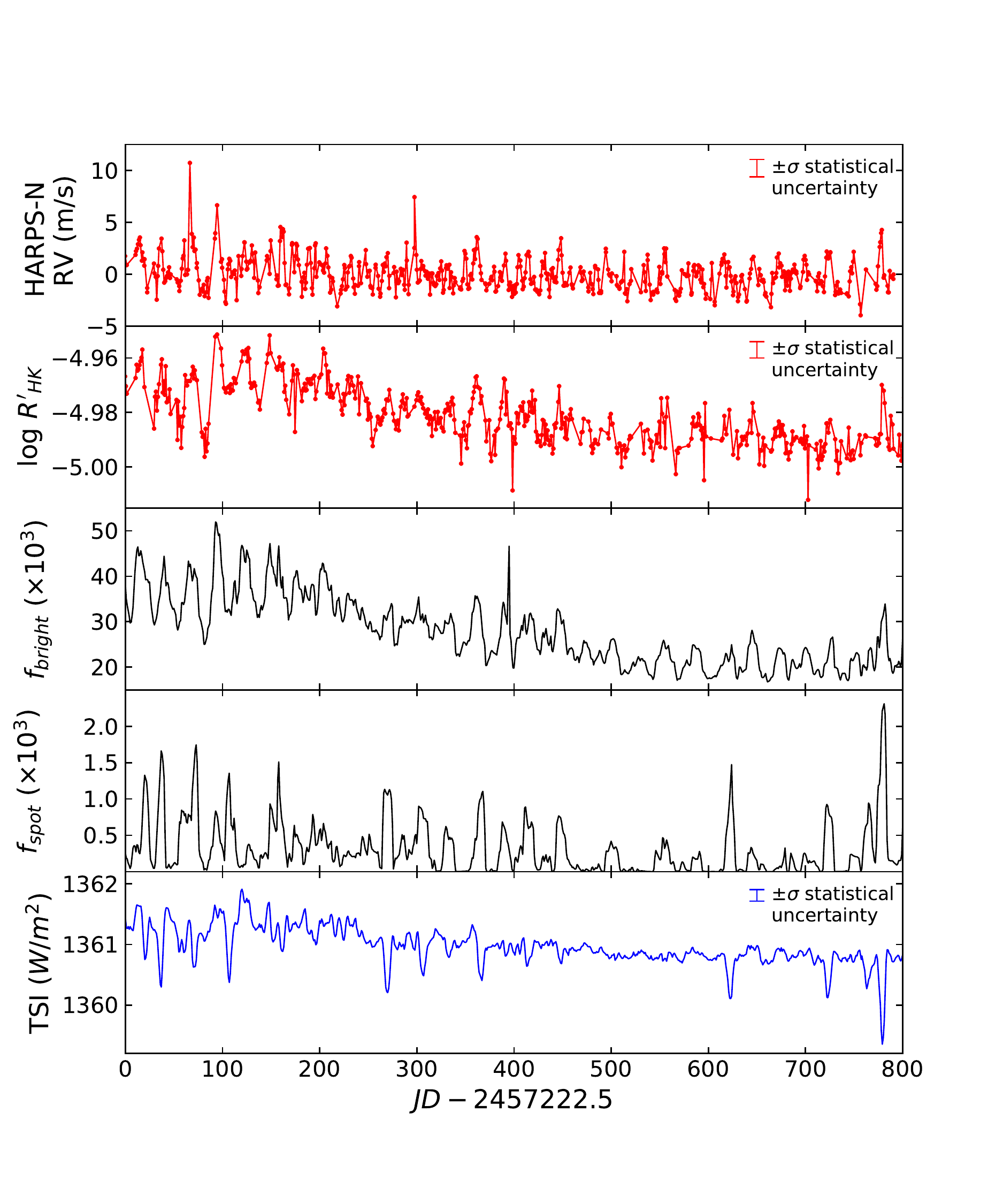} 
\caption{\label{fig:AllVars}Solar measures used in this work. From top to bottom: solar telescope/HARPS-N RV after subtracting the effects due to all planets using the JPL \emph{Horizons} System (measured relative to the averaged HARPS-N solar RV) and $R'_{\rm HK}$ (red), SDO/HMI bright (plage and network) and dark (spot) filling factors (black), and SORCE/TIM TSI (blue). A noticeable decrease in solar activity beginning around Day 200 is visible in all of the displayed activity indicators but not in the HARPS-N RVs. However, we do note an apparent decrease in the RV scatter at this time. Dips in the TSI are coincident with peaks in the spot filling factor. Observations are taken between July 2015 through September 2017, with solar minimum expected in late 2018/early 2019. For the solar telescope/HARPS-N and SORCE/TIM derived quantities, we plot a representative $\pm\sigma$ statistical error bar. Since the SDO/HMI-derived quantities are determined by averaging over $\sim 10^{6}$ CCD pixels, the associated statistical errors are vanishingly small. We therefore omit error bars for those quantities.}
\end{center} 
\end{figure*}

\subsection{Solar Telescope at HARPS-N}

The HARPS-N spectrograph at the TNG is a cross-dispersed echelle spectrograph spanning the visible range \citep{HARPSN_2012}. During the day, a custom-built solar telescope connected to HARPS-N provides a near-continuous stream of disk-integrated solar spectra (\citealt{Xavier_2015}, \citealt{Phillips2016}). This instrument, in operation since 2015, works in combination with the HARPS-N spectrograph to observe the Sun as a star, giving unprecedented temporal coverage (about one exposure every five minutes, with a typical daily coverage of 6 hours) of the solar spectrum with resolving power $R = 115000$ and optical bandwidth spanning 383 nm - 690 nm. The solar telescope has a 3" lens that feeds an integrating sphere, which scrambles all angular information and converts solar images into the equivalent of a point source. Systematic laboratory and on-sky tests show the solar telescope captures the full disk of the Sun with RV precision below 10 cm~s$^{-1}$ as compared to independent SDO/HMI images, well below the 40 cm~s$^{-1}$ per exposure precision of HARPS-N itself (\citealt{HARPS-N_2014}, \citealt{Phillips2016}).

We acquire five minute solar exposures to average over solar acoustic oscillations (\emph{p}-modes), and achieve RV precision of approximately 40 cm~s$^{-1}$. This temporal coverage allows us to investigate solar activity on timescales between minutes and years, as demonstrated in \citealt{Phillips2016}. To reduce the effects of solar oscillations, granulation, and other processes with variability timescales less than 24 hours, we take daily averages of the solar RVs. The changes in these daily-averaged RVs are therefore dominated by stellar activity effects (\citealt{Xavier_2015}, \citealt{Meunier_et_al_2010}).

 Solar RVs are derived from the measured spectra using the HARPS-N Data Reduction System (DRS) (\citealt{Baranne_et_al_1996}, \citealt{Sosnowska_2012}). The contributions of planetary reflex motion to the solar RVs are removed using JPL Horizons ephemerides to determine the Sun-TNG relative velocity \citep{Horizons_1996}. The effects of differential atmospheric extinction are removed from the RVs by calculating the intensity-weighted mean rotational velocity across the solar disk, accounting for the extinction gradient across the disk and the inclination of the solar rotation axis to the local vertical, as described in \cite{CollierCameron_2018}. Exposures contaminated by clouds are identified using the HARPS-N exposure meter, and are removed from the final data set. If any of the 1-second sampled exposure meter measurements are below a certain threshold, the corresponding exposure is rejected. The remaining "RV residuals" are predominantly  the result of solar variability: i.e., if the Sun were a uniform, homogeneous disk, they would be consistent with zero and limited to statistical noise and residual spectrograph systematic variations. These residuals have an RMS amplitude of 1.6 m~s$^{-1}$, comparable to those observed on stars of similar activity levels \citep{Isaacson_Fischer_2010}. Additionally, we extract the calcium S-index, a known correlate of magnetic activity and the derivative $R'_{\rm HK}$ (\citealt{Vaughan_1978}, \citealt{Noyes_et_al_1984}), from the Ca II H\&K lines in the solar spectra. The resulting values of the HARPS-N RVs and $\log{R'_{\rm HK}}$ are shown in the top two panels of Fig.~\ref{fig:AllVars}.

\subsection{SDO/HMI}
HMI onboard SDO captures full disk images of the Sun with near single-granule resolution (\citealt{Schou2012}, \citealt{Pesnell2012}). HMI determines the continuum intensity, line depth, line width, doppler velocity, and magnetic flux at each point along the solar disk by measuring six wavelengths around the 6173.3 \AA\ neutral iron (Fe I) line in two polarization states \citep{Couvidat2016}.

Using thresholding algorithms pioneered by \cite{Fligge_et_al_2000} and subsequently used for solar RV modelling by \cite{Meunier_et_al_2010_MDI} and \cite{Haywood_et_al_2016}, we identify active regions along the solar disk and calculate the magnetic filling factor, $f_{total}$. (See Fig.\ref{fig:AllVars}), the percentage of the solar disk covered by magnetic activity. We use the same intensity thresholds determined by \cite{Yeo_et_al_2013} and employed by \cite{Haywood_et_al_2016} to distinguish between dark regions (sunspots) and bright regions (plage and network), allowing us to calculate filling factors for each type of magnetic feature ($f_{bright}$ and $f_{spot}$ respectively). By combining the intensity and magnetic flux information with the Doppler velocities, we estimate the contributions of magnetic activity to solar RVs.

In this work, we consider the 720 second exposure line-of-sight measurements of the continuum intensity, magnetic field, and Doppler velocity\footnote{Publicly available at \href{url}{http://jsoc.stanford.edu/}}. We use six images each day, sampled evenly over the 2.5 year operational period of the solar telescope at HARPS-N. Note that all HMI observables have a strong 24-hour modulation related to an imperfect removal of the SDO spacecraft's orbit \citep{Couvidat2016}; to mitigate the effects of these and other systematics (\citealt{LB_S_2013}, \citealt{Reiners2016}, \citealt{Hoeksema2018}) we therefore reference all derived RVs to the quiet-sun velocity and take daily averages of the derived filling factors and activity-driven RVs (\citealt{Meunier_et_al_2010_MDI}, \citealt{Haywood_et_al_2016}). (See Sec.~\ref{EstimatingActivity} and Appendix \ref{appA} for further discussion of these calculations.) 

\subsection{SORCE/TIM}
TIM onboard SORCE measures the TSI using a set of four Electrical Substitution Radiometers, providing a near-continuous stream of solar photometry analogous to \emph{Kepler} data \citep{Kopp2005b} for the Sun\footnote{Publicly available at \href{url}{http://lasp.colorado.edu/home/sorce/data/tsi-data/}}. On timescales between days and months, changes in the TSI are related to the movement of bright and dark active regions across the solar surface (as shown in Fig. \ref{fig:AllVars}). The TSI therefore functions as a purely photometric measure of the solar activity. While the solar telescope at HARPS-N is equipped with an exposure meter, variable atmospheric transparency, the aging of telescope components, and the lack of a reference source makes ground-based photometry impractical. The space-based SORCE/TIM is therefore a valuable tool in the study of solar activity, allowing us to compare our disk-integrated and disk-resolved data products with simultaneous photometry.
%

\section{Comparing Measurements of Solar Magnetic Activity}
\label{ActivityComparisions}
The solar telescope, SDO/HMI, and SORCE/TIM each provide a unique lens for analyzing solar activity. Using the broadband, spectroscopic information derived from the solar telescope/HARPS-N, we extract the solar RVs, Mt. Wilson S-index and the derivative index $\log{R'_{HK}}$ (see below for a further discussion of these activity indicators). SDO/HMI directly images active regions on the solar disk, allowing us to identify them as sunspots or plage and network. SORCE/TIM measurements of the TSI provide a photometric measurement of solar activity.

Comparing the time series of each activity indicator, shown in Fig.\ \ref{fig:AllVars}, we see qualitative agreement between the data products of each instrument. The HARPS-N-derived $\log{R'_{\rm HK}}$, the SDO/HMI bright region filling factor (plage and network), the peak amplitudes of the SDO/HMI spot filling factor, and the SORCE/TIM TSI all show the same downward trend as the Sun approaches solar minimum. Furthermore, peaks in the spot filling factor are coincident with sharp dips in the SORCE/TIM TSI. In this section, we make quantitative comparisons between these independent measurements of solar activity and demonstrate the instruments provide a consistent picture of solar magnetic processes. 

\subsection{Comparison of SDO/HMI with Solar Telescope/HARPS-N}
Magnetic heating of the solar chromosphere results in enhanced emission reversals in the cores of the Ca II H\&K lines \citep{Linksy_Avrett_1970}. The observed correlation between these emission reversals and sunspot number, as described in \cite{Wilson_1968}, led to the development of the Mt. Wilson S-index and the color-corrected $\log{R'_{HK}}$, as defined in \cite{Vaughan_1978}. Given the correlations between chromospheric plage and the photospheric spots and faculae, we expect a high degree of correlation between $\log{R'_{HK}}$ and the magnetic filling factors as well \citep{Shapiro_et_al_2014}.

\begin{figure*} 
\begin{center} 
\includegraphics[width=.95\textwidth]{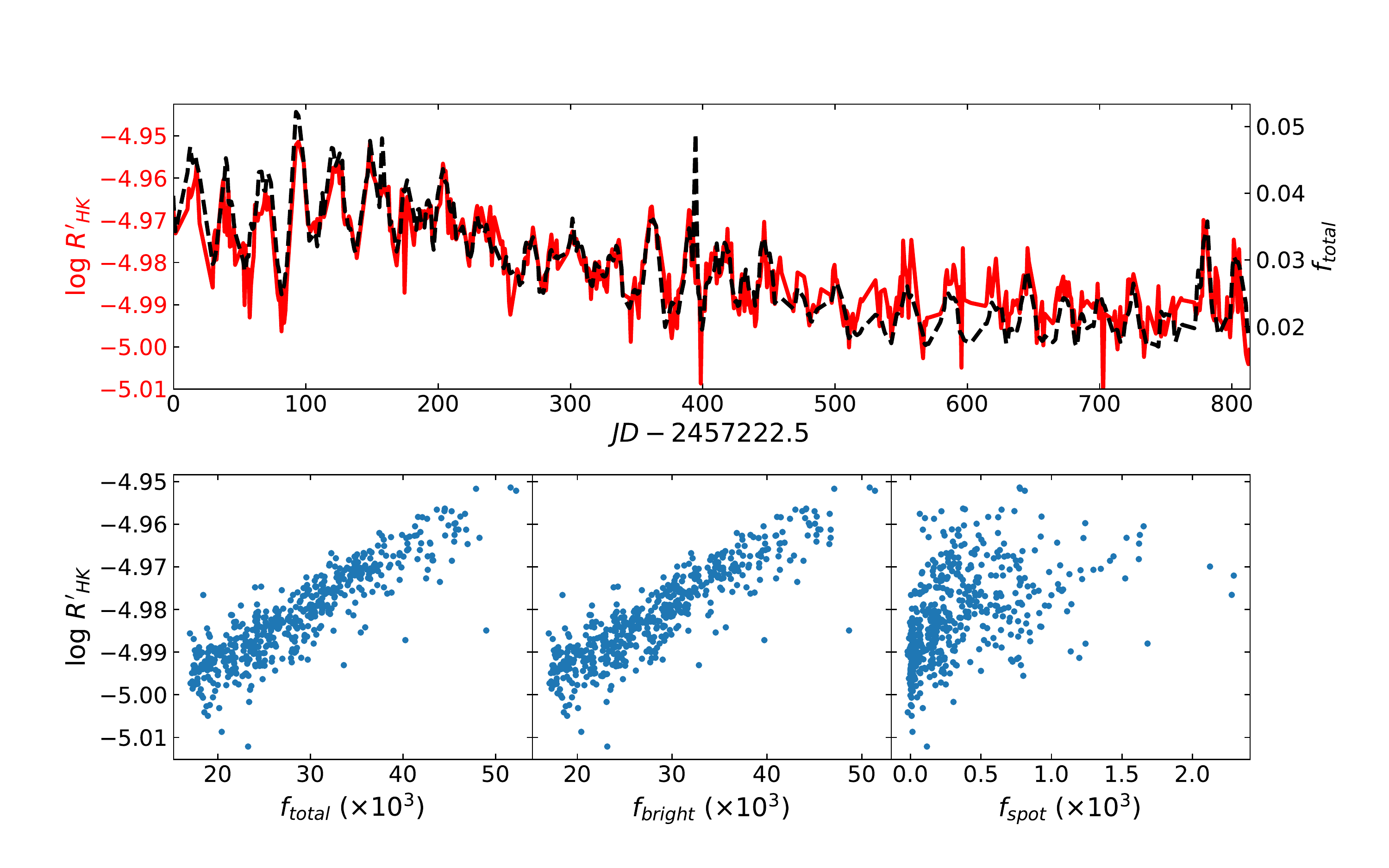} 
\caption{\label{fig:ActFracsInds} \emph{Top:} Spectrally derived $\log{R'_{HK}}$ (black dotted line) and SDO/HMI-calculated total magnetic filling factor (red solid line), plotted as a function of time. A strong correlation between the two quantities is clearly visible in the time series. Both indicators demonstrate oscillations at the synodic solar rotation period (28 days). \emph{Bottom:} Correlation plots between $\log{R'_{HK}}$ and the total filling factor (left), the network and plage filling factor (center), and the spot filling factor (right). We see that the correlation between between the filling factor and $\log{R'_{HK}}$ is driven by the bright regions: the Sun is a plage-dominated star entering activity minimum, resulting in a factor of $\sim 10^2$ fewer sunspots, and a much weaker correlation with the spot filling factor. This relationship is captured by the Spearman correlation coefficients for each filling factor and $\log{R'_{HK}}$: the correlation coefficient between the overall filling factor and $\log{R'_{HK}}$ is 0.8836, the correlation coefficient between the plage/network filling factor and $\log{R'_{HK}}$ is 0.8833, and the correlation coefficient between the spot filling factor and $\log{R'_{HK}}$ is 0.590.}
\end{center} 
\end{figure*}

The SDO/HMI-derived magnetic filling factor and spectroscopic measurements of $\log{R'_{HK}}$ are highly correlated, with a Spearman correlation coefficient of 0.8836. (Fig.~\ref{fig:ActFracsInds}). Dividing the total magnetic filling factor into a bright (plage and network) and dark (spot) contributions also shows the expected behavior; the plage and network, which dominate the total activity, are strongly correlated with $\log{R'_{HK}}$. The spots, however, cover a much smaller portion of the solar surface ($f_{bright}/f_{spot}>80$) and exhibit a much weaker correlation with $\log{R'_{HK}}$.

\subsection{Comparison of SDO/HMI with SORCE/TIM TSI}\label{SORCEresults}
The presence of bright (plage and network) and dark (sunspots) features on the solar surface causes the TSI to fluctuate in response to magnetic activity. This is readily apparent in Fig.\ \ref{fig:AllVars}:  spikes in the spot filling factor derived from SDO/HMI are accompanied by corresponding decreases in the SORCE/TIM TSI and the long-term decrease in the plage and network filling factor are correlated with the long-term decrease of the TSI.

The different brightnesses of these features arise because bright plage and network regions are hotter than the quiet Sun, and that spots are colder. We therefore define $\Delta T_{bright}$ and $\Delta T_{spot}$, the brightness temperature contrasts of these two features. Following \cite{Meunier_et_al_2010}, we use the SDO/HMI derived plage and spot filling factors to reproduce the measured TSI:

\begin{multline}
\label{TSImodel}
TSI = \mathcal{A} \sigma[(1-a_{spot} f_{spot}-a_{bright}f_{bright})T_{quiet}^4 \\+ a_{spot}f_{spot}(T_{quiet} + \Delta T_{spot})^4 \\+ a_{bright}f_{bright}(T_{quiet}+\Delta T_{bright})^4],
\end{multline}

where $\sigma$ is the Stefan-Boltzmann constant, $\mathcal{A}= (R_{\odot}/1 AU)^2$ is a geometrical constant relating the energy emitted at the solar surface to the energy received at Earth, $T_{quiet}$ is the quiet Sun temperature, and $f_{spot}$ and $f_{bright}$ are the HMI spot and plage/network filling factors.

Additionally, $a_{bright}$ and $a_{spot}$ are scaling factors, used to account for systematic differences in the calculation of filling factors. The values of the bright and spot filling factors depend strongly the choice of magnetic flux and intensity thresholds used to differentiate spots, plage, and quiet sun, as well as the  wavelength(s) used to observe these features. Variations in these parameters mean that established sunspot datasets may differ by over 50\% \citep{Meunier_et_al_2010}. Including these scaling factors in our model allows us to account for these definition-dependent factors and to compare the brightness temperature contrasts of each feature to literature values. Since these scaling factors are constant, multiplicative values, they do not affect correlations between the filling factors and the other activity measurements.

A wide range of spot and plage/network temperature contrasts are given in the literature. From \cite{Meunier_et_al_2010}, we infer that $-649 \rm \ K < \Delta T_{spot}< -450 \rm \ K $ and $38 \rm \ K < \Delta T_{bright}< 55 \rm \ K $. Note that the apparent temperature of plage varies with position on the solar disk: Since the disk-averaged SDO/HMI plage filling factor contains no spatial information, we take $\Delta T_{bright}$ to be the average brightness temperature contrast of solar plage. In our analysis, we assume $\Delta T_{spot} = -550 \rm K$ and $\Delta T_{bright} = 46.5 \rm K$, corresponding to the midpoints of the above ranges.

\begin{figure*} 
\begin{center} 
\includegraphics[width=0.95\textwidth]{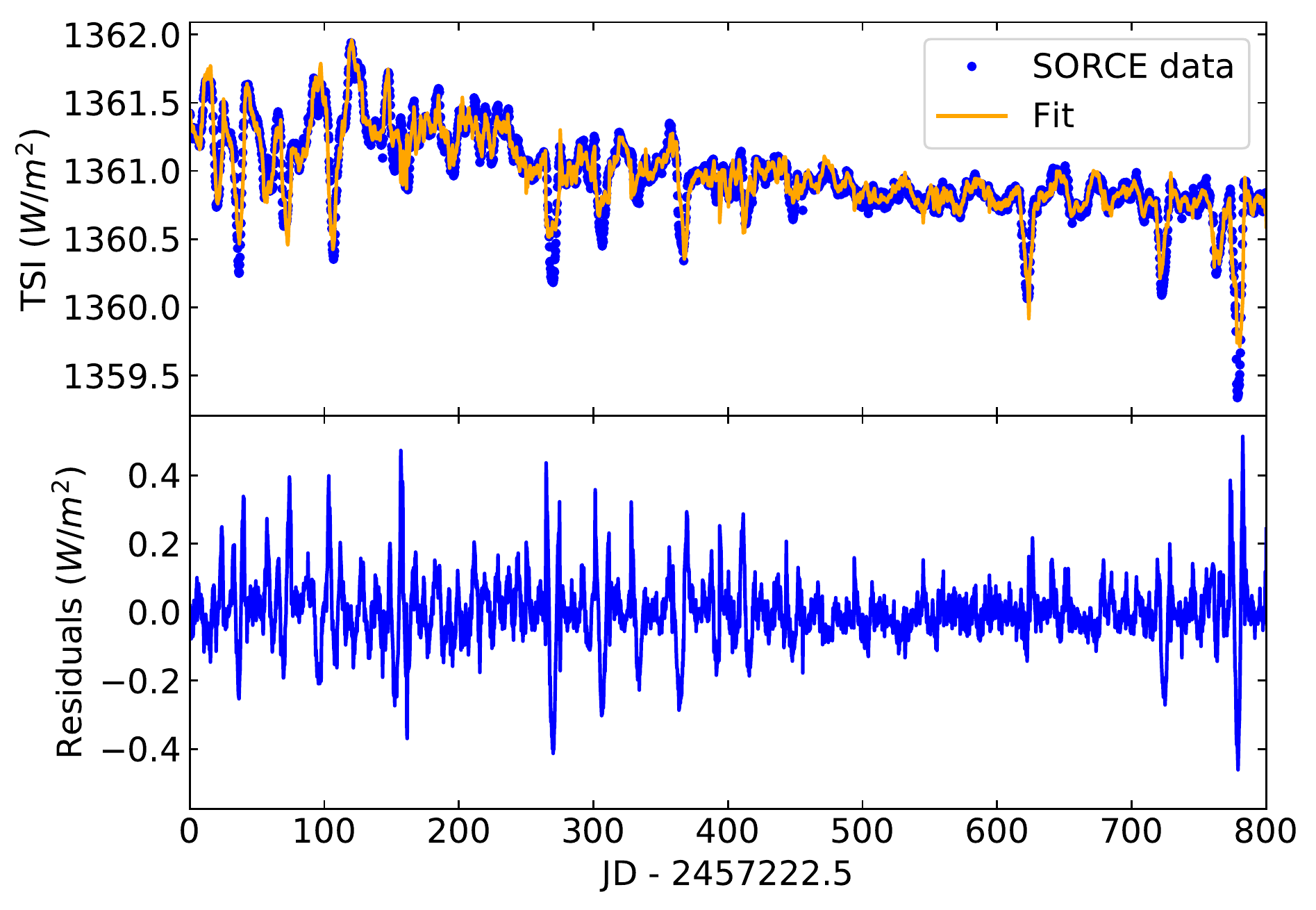} 
\caption{\label{fig:TSI} \emph{Top}: SORCE/TIM measurements of TSI (blue circles), with reconstructed TSI from SDO/HMI filling factors (orange line). \emph{Bottom}: Fit residuals. The residuals shown are consistent with the typical SORCE/TIM uncertainty per data point, 0.48 W~m$^{-2}$. Note that the correlation between the TSI and the convective magnetic shift (see Sec.\ \ref{VconSec}) implies an RV scaling with TSI of 3.3~(m~s$^{-1}$) / (W~m$^{-2}$).}
\end{center} 
\end{figure*}

Assuming the above values of $\Delta T_{spot}$ and $\Delta T_{bright}$ and  using the SDO/HMI-derived time series of $f_{spot}$ and $f_{bright}$, we fit Eq.~\ref{TSImodel} to the SORCE/TIM TSI as shown in Fig.~\ref{fig:TSI}. From this fit, we extract a quiet-Sun temperature ($T_{quiet} = 5770.080 \pm 0.007 \rm K$) and scaling coefficients ($a_{bright} = 0.959 \pm 0.007$ and $a_{spot} = 1.54 \pm 0.02$). These coefficients are consistent with the 20\% and 50\% definitional variations in feature area described in \cite{Meunier_et_al_2010}.

The three-way agreement between the solar telescope $\log{R'_{\rm HK}}$, SDO/HMI filling factors, and SORCE/TIM TSI indicates that our activity models provide a consistent picture of solar magnetic processes. However, the solar telescope/HARPS-N solar RVs are not in full agreement with these activity measurements. In particular, Fig. \ref{fig:AllVars} shows that $\log{R'_{\rm HK}}$, $f_{bright}$, and the TSI all display a downward trend over the 800 day observation period. The solar RVs do not display this trend. To quantify this disagreement, we compute the Spearman correlation coefficient between the solar RVs and $\log{R'_{\rm HK}}$, $f_{bright}$, and TSI, yielding values of 0.42, 0.40, and 0.04 respectively. To understand this discrepancy, we now use SDO/HMI-derived RVs to reproduce solar telescope/HARPS-N measurements.
\newline

\section{Calculating RV Contributions of Spots/plage} \label{EstimatingActivity}

We model the effects of stellar magnetic activity on RV measurements as a combination of two processes: the suppression of convection in magnetically active regions that leads to a net redshift of the spectrum ($\Delta \hat{v}_{\rm conv}$), and the effect of bright and dark active regions on the solar disk that leads to a photometric shift ($\Delta \hat{v}_{\rm phot}$). In the following sections, we discuss the physical origins of each term and their magnitudes as derived from SDO/HMI images. We then reconstruct the solar RVs  from a combination of the two processes and fit this model to the RVs measured with the solar telescope/HARPS-N.

\subsection{Suppression of Convective Blueshift, $\Delta \hat{v}_{\rm conv}$}
\label{VconSec}
$\Delta \hat{v}_{\rm conv}$ results from the suppression of solar convective motions by local magnetic fields. Taking an intensity-weighted average of the bright, upflowing plasma in the middle of the convective cells and dark, downflowing plasma at the cell edges results in an overall convective blueshift with an amplitude of approximately 250 m~s$^{-1}$ (\citealt{Dravins_1981}, \citealt{Meunier_et_al_2010}). The plasma's interaction with solar magnetic fields impedes this convective motion and therefore attenuates this convective blueshift. Note that the convective blueshift of an observed spectral line depends on its formation depth in the photosphere (\citealt{Gray2009}, \citealt{GrayOostra2018}). The convective shift observed by SDO/HMI using the 6173.3 \AA\ line will therefore differ from the solar telescope/HARPS-N observations, which are averaged over many lines. To account for this systematic difference, we apply a scaling coefficient in our RV reconstruction as discussed in Sec. \ref{RVRecon}.

In previous studies, \cite{Meunier_et_al_2010_MDI}, \cite{Dumusque_et_al_2014}, and \cite{Haywood_et_al_2016} found $\Delta \hat{v}_{\rm conv}$ to be the dominant source of RV shifts, with a disk integrated amplitude of several m~s$^{-1}$. Using the SDO/HMI dopplergrams in conjunction with the magnetic flux and continuum intensity images, we replicate the analysis of \cite{Haywood_et_al_2016} to determine $\Delta \hat{v}_{\rm conv}$ for the full solar telescope/HARPS-N observing period. Several m~s$^{-1}$ variations are observed at the synodic rotation period of the Sun along with long-term drifts of a similar amplitude as shown in the upper-left panel of Fig.~\ref{fig:FitBasis}.

\subsection{Photometric Shift, $\Delta \hat{v}_{\rm phot}$}

\begin{figure*} 
\begin{center} 
\includegraphics[width=.95\textwidth]{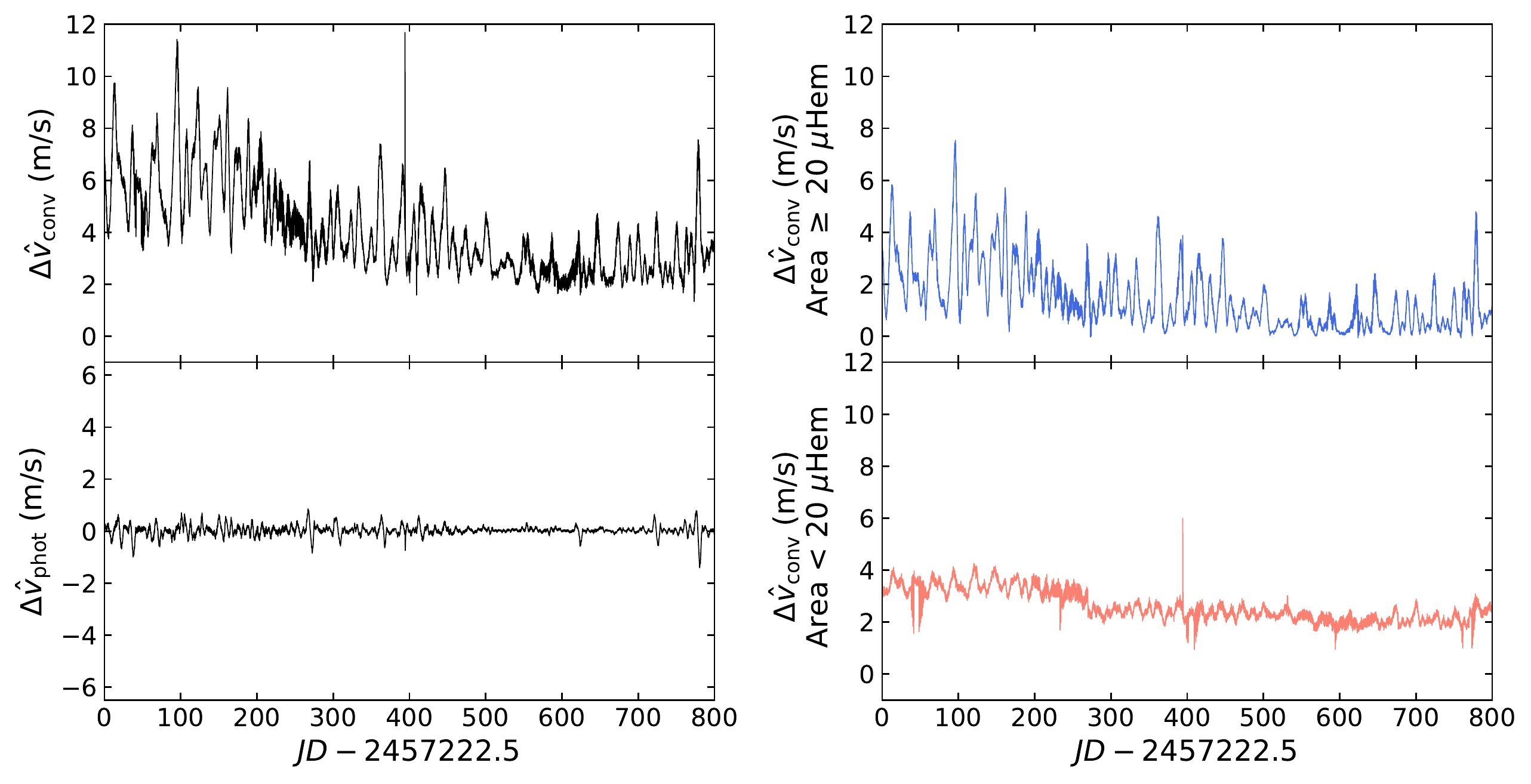} \caption{\label{fig:FitBasis} \emph{Left:} SDO/HMI-derived estimates of the convective (\emph{top}) and photometric velocities (\emph{bottom}) using all active regions. \emph{Right:} Contributions of plage (area $\geq 20$ $\mu\rm{Hem}$, \emph{top})  and network (area $< 20$ $\mu\rm{Hem}$, \emph{bottom}) to the suppression of convective blueshift. As in previous works (\citealt{Meunier_et_al_2010_MDI}, \citealt{Haywood_et_al_2016}), we find that $\Delta \hat{v}_{\rm conv}$ dominates the effects of $\Delta \hat{v}_{\rm phot}$.}
\end{center} 
\end{figure*}

The presence of dark sunspots and bright plage on the solar disk break the Sun's symmetry about its rotation axis. This results in an imbalanced Doppler shift across the solar disk; it is this Doppler imbalance that results in $\Delta \hat{v}_{\rm phot}$, the photometric RV shift due to magnetic activity (\citealt{Saar_et_al_1997}, \citealt{Lagrange_et_al_2010}). As before, we use the methods of \cite{Haywood_et_al_2016} to compute $\Delta \hat{v}_{\rm phot}$ using the SDO/HMI-measured full-disk magnetograms and continuum intensity. This time series is also shown in the lower-left panel of Fig. \ref{fig:FitBasis} and, as expected, is significantly smaller than the shifts calculated for $\Delta \hat{v}_{\rm conv}$.

\subsection{Reconstruction of Solar RVs From SDO/HMI Basis Functions}\label{RVRecon}

Following \cite{Haywood_et_al_2016}, we  model the total solar telescope RVs, $\Delta \rm RV_{\rm model}$, using a linear combination of $\Delta \hat{v}_{\rm conv}$ and $\Delta \hat{v}_{\rm phot}$:
\begin{equation}
\Delta {\rm RV}_{\rm model} = A(t) \Delta \hat{v}_{\rm phot} + B(t) \Delta \hat{v}_{\rm conv} + {\rm RV}_{0}\label{HaywoodModel}.
\end{equation}
Here $A(t)$ and $B(t)$ are weighting coefficients for the photometric and convective RV shifts, and $RV_{0}$ describes the zero point of HARPS-N. As $RV_{0}$ is a purely instrumental parameter, we expect it to remain constant with time. $A(t)$ and $B(t)$ describe the mapping of information from the single $\lambda = 6173.3$ \AA\  spectral line onto the several thousand lines used in the HARPS-N CCF analysis (\citealt{Baranne_et_al_1996}, \citealt{Sosnowska_2012}). The coefficient $A(t)$ accounts for systematic differences between the bright and dark active regions observed with SDO/HMI and the spectrum observed with the solar telescope/HARPS-N, analogous to the scaling factors used in our TSI reconstruction (see Eq. \ref{TSImodel}). The coefficient $B(t)$ accounts for the systematic difference in the convective blueshift due to the different heights of formation of each spectral line. We thus expect $A(t)$ and $B(t)$ to be of order unity, but not necessarily equal to 1. In \cite{Haywood_et_al_2016}, $A(t)$, $B(t)$, and $RV_0$ are taken to be constant. However, these coefficients could vary with time perhaps due to additional magnetic processes at work, or some other changes over the activity cycle. We divide each time series into smaller subsections and calculate the fits for each subsection to investigate how $A(t)$ and $B(t)$ evolve in time.

As discussed in Sec. \ref{sec:Obs}, we take daily averages of 2.5 years of solar telescope data to mitigate the effects of solar $p$-modes and granulation. We expect that, on timescales longer than several days, the measured RV variations are dominated by magnetic effects. We then fit Eq.\ \ref{HaywoodModel} to the whole data set, yielding global values of $A(t)$ and $B(t)$, as given in Table \ref{tab:GlobFitTable}.
Following \cite{Haywood_et_al_2016}, we include an uncorrelated noise parameter $s$, added in quadrature to the solar telescope/HARPS-N observational errors, to account for instrumental uncertainties and other processes not in our model \citep{ACC_et_al_2006}.

We then divide the data into subsections of $N = $ 112 days (corresponding to four synodic solar rotations per subsection) and repeat the fit, evaluating $A(t)$, $B(t)$, and $RV_0$ for each subsection. We chose this value of $N$  to maximize the number of data sections while maintaining sufficiently small statistical uncertainties. The results described below do not depend on the exact value of $N$.

\begin{figure*} 
\begin{center} 
\includegraphics[width=.95\textwidth]{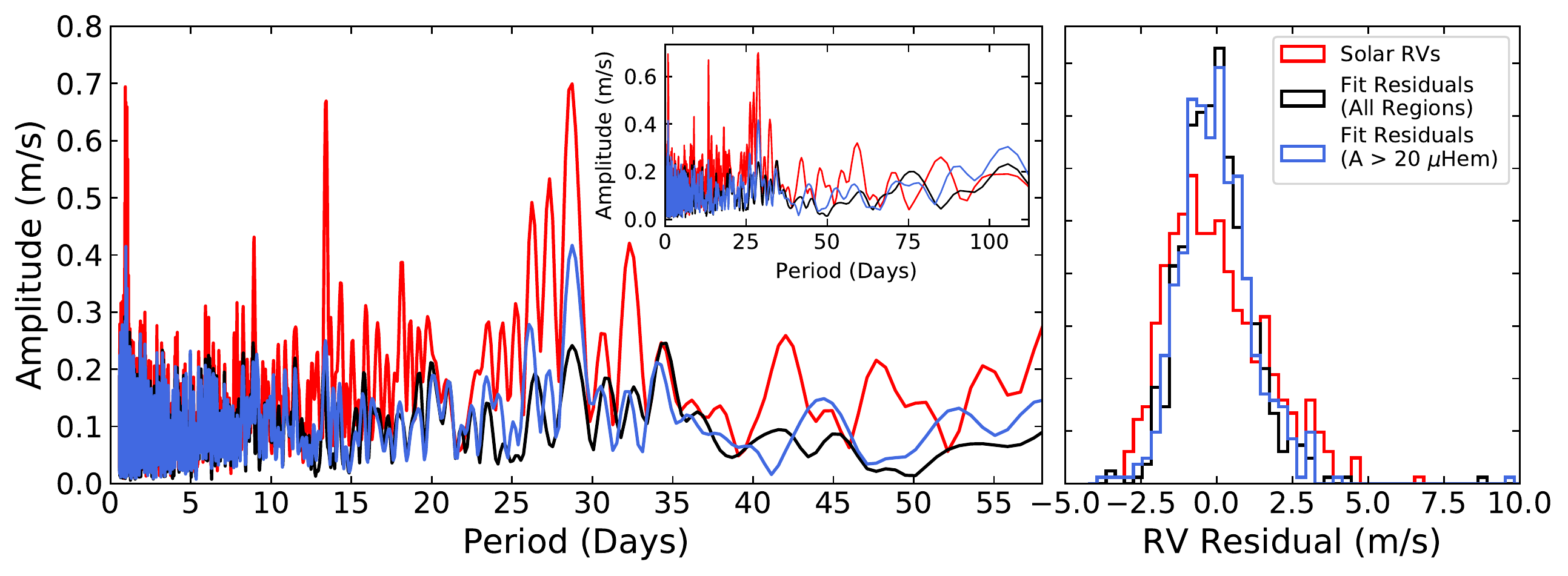} \caption{\label{fig:FitResPerdHist} \emph{Left:} Periodogram of the solar telescope RVs (red), fit residuals using all active regions (black), and fit residuals using a 20 $\mu\rm{Hem}$ area threshold (blue). We note that measured solar RVs have an amplitude of 0.72 m~s$^{-1}$ at the solar rotation period. Applying our model with no area cut reduces this amplitude to 0.24 m~s$^{-1}$; including an area cut results in an amplitude of 0.42 m~s$^{-1}$. 
\emph{Inset:} A zoomed-out view of the periodgram. We note that the two fits successfully reduce the RV amplitudes observed on most timescales greater than the rotation period. \emph{Right:} Histogram of the RV residuals. Both fits result in Gaussian-distributed RV residuals: while both fits display decrease the RMS RV residuals, applying an area threshold does not produce a visible change in the fit residuals. The area cut does, however, remove the unphysical trend in $RV_0$, as discussed in Sec. \ref{sec:DeltaRV0Discussion} and as shown in Fig. \ref{fig:FitResults}}.
\end{center} 
\end{figure*}

Since the RV contributions of magnetic active regions are modulated by the Sun's rotation (see Figs. \ref{fig:FitBasis} and \ref{fig:vcon_v_area}), we expect our model to fully capture RV variation on timescales of the rotation period (and its harmonics) and above. As shown in Fig. \ref{fig:FitResPerdHist}, the dominant contributions to the observed RV variations occur on these timescales. Below the rotation period, the solar RV is modulated to some degree by magnetic region growth and decay, but also by granulation and supergranulation due to surface magneto-convection. Our model is not designed to capture these convective processes, and we therefore do not expect it to capture all RV variations on these short timescales.

\subsection{Active Region Area Dependence of Convective Shift}\label{sec:AreaDependence}

In addition to reconstructing the solar telescope/HARPS-N RVs variations we investigate if and how the suppression of convective blueshift associated with a given active region depends on its size. \cite{Meunier_et_al_2010} speculated that small intergranular network features and large plage/sunspot regions would have different contributions to the convective blueshift, and \citeauthor{Palumbo2017} (\citeyear{Palumbo2017}; \citeyear{Palumbo2019}), observed different center-to-limb velocity variations for solar network and plage. We differentiate the RV contributions of these two classes of active region. The network and plage/spot regions may be distinguished based on their spatial distributions: while small network are uniformly distributed over the solar disk, large plage/spot regions appear only around active latitudes, leading to the well-known butterfly diagram (see \cite{Hathaway2015} and references there-in).

\begin{figure*} 
\begin{center} 
\includegraphics[width=0.95\textwidth]{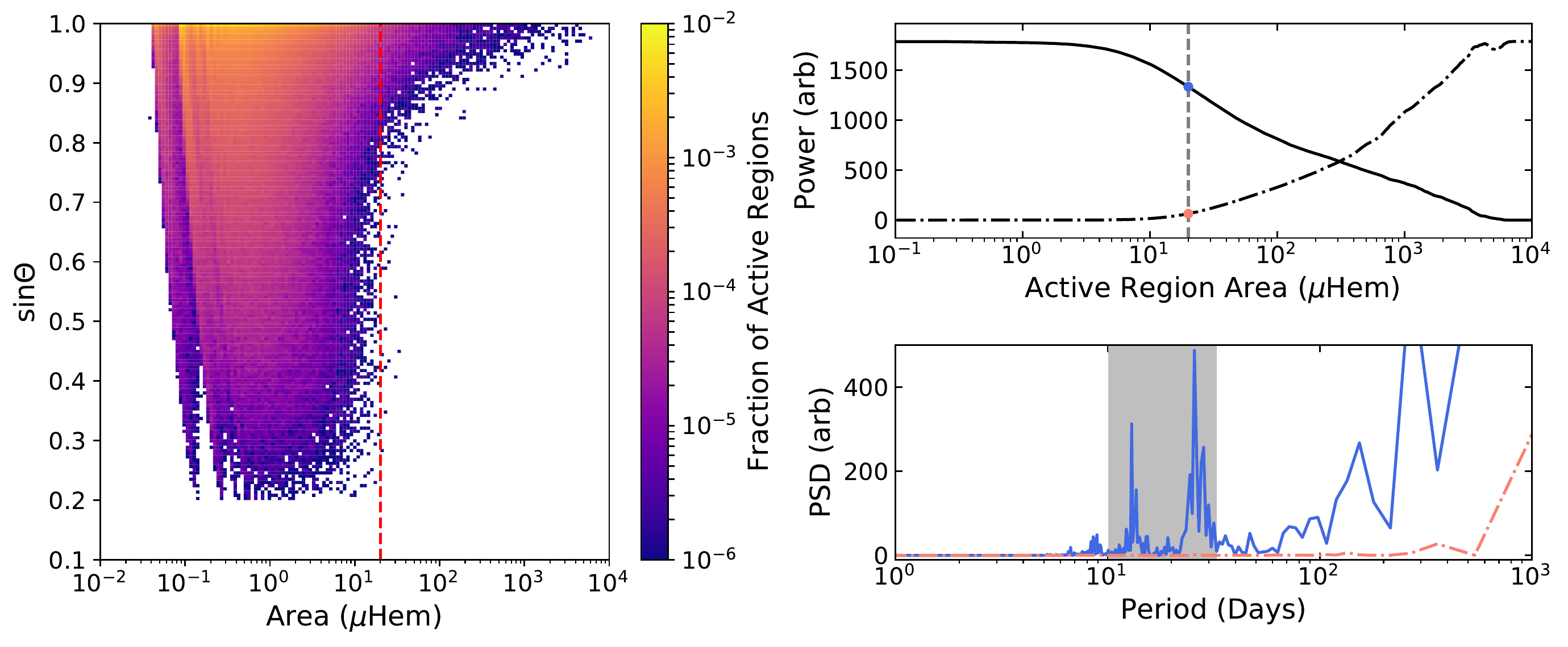}
\caption{\label{fig:vcon_v_area} \emph{Left}: Fraction of observed solar active regions as a function of region area and co-latitude, $\Theta$ (as measured from the north pole). From the observed spatial distribution, we may divide the active regions into small network, which appear all across the solar disk, and large spots/plage, which preferentially appear around activity latitudes. The sharp change in the spatial distribution allows us to infer the presence of a sharp cut-off allows us to infer an area threshold of 20 micro-hemispheres that separates these two regimes.
\emph{Upper-right}: Power associated with the solar rotation period and its first harmonic above (solid) and below the area threshold (dashed). Power at these frequencies is evaluated by integrating the power spectral density (PSD) of the RV contributions for each region size over the shaded region, indicated below. \emph{Lower-right}: PSD of RV contributions above (solid blue) and below (red dashed)  micro-hemisphere. Below the 20 micro-hemisphere threshold, there is little power associated with the solar rotation period: these small structures therefore do not contribute to the solar RVs on the timescales of interest in this work.}
\end{center} 
\end{figure*}

To distinguish network from plage/spot regions, we plot the 2D distribution of active region co-latitude $\Theta$ and area, as shown in the left panel of Fig. \ref{fig:vcon_v_area}. There is a sharp cut at approximately 20 micro-hemispheres (that is, 20 parts per million of the visible hemisphere), or $60 \ \rm Mm^2$ in areal units. Active regions smaller than this cutoff are distributed across the Sun, while regions above the cutoff only appear around the equator, at $0.75 < \sin{\Theta} \leq 1$. We therefore use this area threshold to classify each active region as small network or large spot/plage.

\begin{figure} 
\begin{center} 
\includegraphics[width=.45\textwidth]{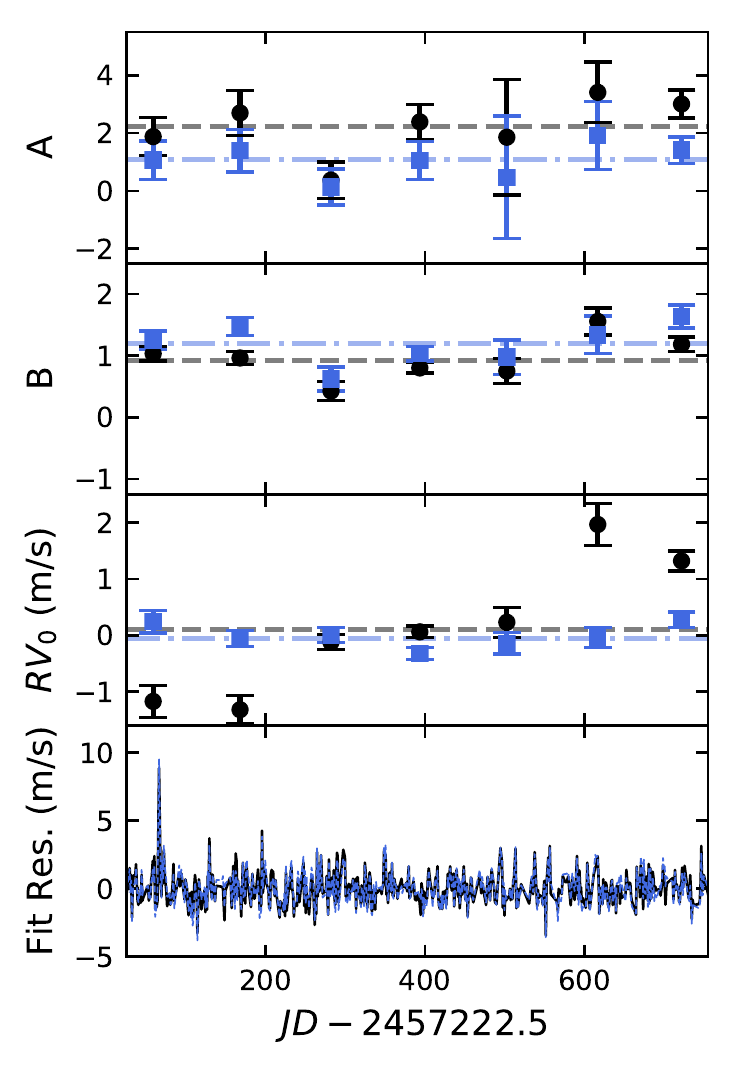} \caption{\label{fig:FitResults} \emph{First and second panel}: Fit parameters $A(t)$, $B(t)$, for the unitless scaling parameters for the photometric and convective RV shifts, $\Delta \hat{v}_{\rm phot}$, and $\Delta \hat{v}_{\rm conv}$ derived from SDO/HMI (plotted in Fig.\ \ref{fig:FitBasis}). \emph{Third panel}: RV offset ($RV_0(t)$) in m~s$^{-1}$. Parameters fitted to HARPS-N solar RVs using Eq.~\ref{HaywoodModel} (\emph{see text}) in $N = $ 112 day sets, using all active regions (black points) or a 20 $\mu\rm{Hem}$ area threshold (blue squares). Statistical error bars are plotted for each  parameter. \emph{Fourth panel}: Residuals of both fits.}
\end{center} 
\end{figure}

To investigate the differing contributions of the network and spot/plage, we compute $\Delta \hat{v}_{\rm conv}$ as a function of time using only network and only spot/plage regions. From the resulting time-series (see Fig. \ref{fig:FitBasis}) and periodograms (right panel of Fig. \ref{fig:vcon_v_area}), we observe that the majority of the RV variability at the solar rotation period is the result of large active regions: small regions do not significantly contribute to the suppression of convective blueshift on this timescale. Given these differing contributions, we perform the RV reconstruction of Sec. \ref{RVRecon} first using the convective RV shift calculated using all observed active regions; second using the convective RV shift calculated using only large spots/plage. The results of this analysis are given in Fig. \ref{fig:FitResults} and Tables \ref{tab:RMSvals} and \ref{tab:GlobFitTable}.

\begin{table*}
\centering
\begin{tabular}{c c c c c}\hline
Basis function & All Active Regions & Area $\geq 20 \mu\rm{Hem}$ & \cite{Haywood_et_al_2016} & \cite{Meunier_et_al_2010_MDI} \\ \hline
$\Delta \hat{v}_{\rm phot}$ & 0.21 m~s$^{-1}$ & 0.21 m~s$^{-1}$  & 0.17 m~s$^{-1}$ &0.42 m~s$^{-1}$\\
$\Delta \hat{v}_{\rm conv}$ & 1.69 m~s$^{-1}$ & 0.88 m~s$^{-1}$ & 1.30 m~s$^{-1}$ & 1.39 m~s$^{-1}$\\
HARPS-N RV & 1.64 m~s$^{-1}$&  &  \\ \hline
\end{tabular}
\caption{\label{tab:RMSvals} RMS amplitudes of RV time series. We include the time series derived using all regions (left column of Fig. \ref{fig:FitBasis}) and using only plage regions (right column of Fig. \ref{fig:FitBasis}). As a point of comparison, we also include the values of \cite{Haywood_et_al_2016} (also derived from SDO/HMI), the values of \cite{Meunier_et_al_2010_MDI} (derived from the Michelson Doppler Imager onboard the Solar and Heliospheric Observatory), and the solar telescope measurements of the solar RVs (top panel of Fig. \ref{fig:AllVars}).}
\end{table*}

\begin{table*}
\centering
\begin{tabular}{c c c c c }\hline
Parameter & Basis function & All Active Regions & Area $ \geq 20 \mu \rm{Hem}$  & \cite{Haywood_et_al_2016}\\ \hline
$A(t)$ & $\Delta \hat{v}_{\rm phot}$ & $2.24 \pm 0.60$ & $1.09 \pm 0.58$  & $2.45 \pm 2.02$\\
$B(t)$ & $\Delta \hat{v}_{\rm conv}$ & $0.93 \pm 0.11$ &$1.20 \pm 0.15 $ & $1.85 \pm 0.27$\\
$RV_0$ &                             & $102.51 \pm 0.06$ m~s$^{-1}$ &$102.36 \pm 0.13$ m~s$^{-1}$ & $99.80 \pm 0.28$ m~s$^{-1}$\\ 
$s$ &                                & 1.21 m~s$^{-1}$& 1.23 m~s$^{-1}$ & $2.70$ m~s$^{-1}$\\ \hline
\end{tabular}
\caption{\label{tab:GlobFitTable} Average values of the SDO/HMI-derived $\Delta \rm RV_{\rm model}$ to solar telescope/HARPS-N RVs using Eq.~\ref{HaywoodModel}. (\emph{See text.}) We provide values derived using both network and plage regions, replicating the analysis of \cite{Haywood_et_al_2016}, as well also values derived using only the plage regions. The time variation of these parameters is shown in Fig. \ref{fig:FitResults}. We also include the results of \cite{Haywood_et_al_2016} as a point of comparison. Error bars on each parameter are statistical uncertainties and $s$ is the added white noise beyond the 40 cm~s$^{-1}$ noise associated with each solar telescope observation. }
\end{table*}

\section{Discussion}

\subsection{Reconstruction of Solar RVs}

As shown in Fig.\ \ref{fig:FitResults}, both $A(t)$ and $B(t)$ are consistent with constant values, implying that $\Delta \hat{v}_{\rm conv}$ and $\Delta \hat{v}_{\rm phot}$ have a consistent effect on the solar telescope/HARPS-N measurements over the full observation period. The resulting $A(t)$ and $B(t)$ values are consistent with those reported by \citeauthor{Haywood_et_al_2016}, as shown in Table \ref{tab:GlobFitTable}. \cite{Haywood_et_al_2016} measured the Sun close to solar maximum, observing a total magnetic filling factor $6\% < f_{total} < 10\% $, whereas near solar minimum, we observe $f_{total} < 5\% $. We also note that the RMS amplitudes of $\Delta \hat{v}_{\rm phot}$ and $\Delta \hat{v}_{\rm conv}$ (shown in Table \ref{tab:RMSvals}), calculated using network and plage regions, are consistent with or somewhat smaller than those of \cite{Meunier_et_al_2010_MDI} and \cite{Haywood_et_al_2016}, which is consistent with observations performed at different parts of the activity cycle. Given the agreement of our $A(t)$ and $B(t)$ values with those of \cite{Haywood_et_al_2016}, we may conclude that these parameters do not change significantly as the Sun enters activity minimum, and only weakly depend on the magnetic filling factor, if they do so at all.

Using only large plage/spot regions in our reconstruction of $\Delta \hat{v}_{\rm conv}$ does not significantly change the magnitude of $A(t)$ and $B(t)$ compared to using all active regions. However, when all active regions are considered, the calculated instrumental offset $RV_0(t)$ (see Fig. \ref{fig:FitResults}) slowly increases over the three year observation period. This slow increase disappears when our model assumes that only large active regions suppress the convective blueshift. We discuss the implications of this result in Sec. \ref{sec:DeltaRV0Discussion}.

\subsection{Long-Timescale Variations: Changes in $RV_0$}
\label{sec:DeltaRV0Discussion}
As our model assumes that magnetic activity is fully described by $\Delta \hat{v}_{\rm conv}$ and $\Delta \hat{v}_{\rm phot}$, we expect $RV_0$ to be an instrument-dependent parameter related to the zero point of HARPS-N, and therefore constant over our observation period. HARPS-N exposures are calibrated using a simultaneous reference with sub-m~s$^{-1}$ precision \citep{HARPS-N_2014}, and the SDO/HMI basis functions are calculated relative to the quiet-Sun velocity \citep{Haywood_et_al_2016}; long-term instrumental drifts are therefore calibrated out of each measurement, and should not affect the value of $RV_0$.

However, in both fits to the solar telescope data, we find a systematic increase in $RV_0$. When all active regions are considered, we obtain a shift of $\Delta RV_0 = 2.6$ m~s$^{-1}$ over the course of the 800 day measurement period. Accounting for the area dependence of the convective velocity eliminates this variation almost entirely. This is consistent with our hypothesis regarding the area dependence of the convective velocity: small active regions do not meaningfully contribute to the suppression of convective blueshift on timescales of the solar rotation period.  Instead, as shown in Fig. \ref{fig:FitBasis} and the right panel of Fig. \ref{fig:vcon_v_area}, these small regions contribute a systematic drift on timescales of hundreds of days. Differentiating the contributions of small and large magnetic activity is therefore necessary for the successful detection of long-period, low mass planets around solar-type stars.

Physically, this systematic RV shift may be due to the different contributions of plage and network, as demonstrated in \citeauthor{Palumbo2017} (\citeyear{Palumbo2017}; \citeyear{Palumbo2019}). Part of the active solar network results from decaying of plage regions, resulting in a correlation between the plage filling factor and the network RV contribution. This, in turn may lead to an overall systematic RV shift. We may also consider a similar scenario for dark spots: Small, dark solar pores lack penumbra, and therefore have a different contribution to the solar RV. As the Sun enters activity minimum, large spots become less common, leading to a larger relative pore contribution, and therefore a systematic RV shift. In both cases, large and small active regions must therefore be treated separately in our RV reconstruction. 

\subsection{RV Residuals and Rotational Modulated Variations}

The measured solar telescope/HARPS-N RV variations (Fig. \ref{fig:AllVars}, top) have an RMS scatter of $1.65$ m~s$^{-1}$. Subtracting the reconstructed values of $\Delta \rm RV_{\rm model}$ computed using all active regions and using the physically-motivated constant value of $RV_0$ (assumed to be an instrumental offset) reduces this scatter to 1.31 m~s$^{-1}$. Repeating this analysis with an empirical time-varying value of $RV_0$ reduces this scatter to 1.18 m~s$^{-1}$. Incorporating the area dependence of the convective shift into our model results in an RMS scatter of $1.21$ m~s$^{-1}$. While including this spatial information does not further improve the RMS scatter of the RV residuals, it almost completely eliminates the observed change in $RV_0$, and thus constitutes a more physically grounded and complete model, as previously discussed in Sec. \ref{sec:DeltaRV0Discussion}.

As discussed at the end of Sec. \ref{RVRecon} and as shown in Fig. \ref{fig:vcon_v_area}, we expect our model to eliminate the observed power at the solar rotation period. The measured solar telescope RV variations have an amplitude of 0.72 $\pm$ 0.06 m~s$^{-1}$ at this timescale: applying our model using all active regions reduces this amplitude to 0.24 $\pm$ 0.08 m~s$^{-1}$. Incorporating the area dependence of the convective shift into our model results in an amplitude of 0.42 $\pm$ 0.08 m~s$^{-1}$ (see Fig. \ref{fig:FitResPerdHist}). The residual signal at the rotation period may indicate that our area threshold does not perfectly differentiate plage and network regions: smaller plage regions are not being included in our calculated value of $\Delta \hat{v}_{\rm conv}$, resulting in an imperfect removal of the RV signal at the rotation period. Reducing the area threshold reduces the residual amplitude at the rotation period: however, the inclusion of network regions in the calculated $\Delta \hat{v}_{\rm conv}$ results in a non-zero trend in $RV_0$. The residual signal may also result from the several-day lag between activity proxies and RV signals observed by \cite{Dumusque_et_al_2014} and \cite{CollierCameron_2018}. Lastly, we note that our model assumes the network regions have the same RV contribution as the quiet Sun: in reality, however, we expect network to provide an additional, nontrivial RV contribution (\citeauthor{Palumbo2017} \citeyear{Palumbo2017}; \citeyear{Palumbo2019}). A more sophisticated model will be required to fully describe these network-driven variations.

In summary, the reconstructed RVs leave over 1 m~s$^{-1}$ of RV variations unaccounted for: these may be the result of supergranulation, which has a physical timescale longer than the 6-8 hour solar observation period at the TNG, (\citealt{DelMoro_et_al_2004}, \citealt{Meunier_et_al_2015}). They may also result from additional surface velocity flows unaccounted for in our model, RV differences of network relative to quiet Sun,  or an unaccounted for instrumental systematic.\footnote{The HARPS/HARPS-N DRS was recently upgraded to improve the stability of the daily wavelength calibrations. At the time of writing, the HARPS-N solar data had not yet been reprocessed with the new DRS. We understand that the older version of the software has an uncertainty of up to 1 m~s$^{-1}$ in the RV zero points between successive days of observation.}

\subsection{Magnetic Activity Indicators and Active Region Area}

The agreement between the magnetic filling factors, $R'_{\rm HK}$, and TSI demonstrated in Sec. \ref{ActivityComparisions} confirms that the traditional metrics for solar activity are all self-consistent. However the correspondence between these activity indicators is not improved by separately considering large and small active regions: the chromospheric emission captured by $\log{R'_{\rm HK}}$ is strongly correlated with the total magnetic filling factor, not the large region filling factor. Similarly, the TSI may only be accurately reproduced using both large and small active regions. All the magnetic active regions on the Sun have an enhanced chromospheric column density that strengthens the emission reversals in the Ca II H\&K line cores, and all bright/dark regions will modulate the Sun's overall brightness. These traditional stellar activity indicators are thus correlated only with the overall coverage of active features, and not the size of each feature. Given the observed dependence of the suppression of convective blueshift on active region size, we may therefore conclude new activity indicators correlated with active region sizes are needed to successfully reproduce RV variations on distant stars.

\section{Conclusions}

In this work, we analyze 3 years of solar observations during the decline of Carrington Cycle 24 to test models of radial-velocity variations of Sun-like stars. We compare solar telescope/HARPS-N measurements of the solar RVs and $\log{R'_{\rm HK}}$, SDO/HMI disk-resolved activity images, and SORCE/TIM measurements of the total solar irradiance. As expected, the observed values of $\log{R'_{\rm HK}}$ and TSI are strongly correlated with the overall magnetic filling factor derived from SDO/HMI images.

However, these activity indicators are not straight-forward predictors of the observed solar RV variations. While we see a slow decrease in $\log{R'_{\rm HK}}$, TSI, and magnetic filling factor as the Sun enters cycle minimum, we do not observe this decrease in the solar telescope/HARPS-N RV variations. To investigate this discrepancy, we model the solar RV as a linear combination of the suppression of convective blueshift and rotational flux imbalance. Our initial reconstruction of the solar RV variations decreased the RMS scatter from $1.65$ m~s$^{-1}$\ to $1.18$ m~s$^{-1}$ and reduced the RV amplitude at the rotation period by a factor of 4, but only by introducing an arbitrary systematic drift of $2.6$ m~s$^{-1}$ over the 800 day observation period. By computing contribution of each active region to the suppression of convective blueshift, we find that active regions smaller than 20 ppm ($60 \ \rm Mm^2$) do not significantly suppress the convective blueshift. Including this area dependence in our model does not further decrease the overall RMS scatter, and results in a factor of 2 reduction of the RV amplitude at the rotation period. However, it completely eliminates the need to introduce an arbitrary systematic drift in our reconstructed RVs, resulting in a more physically-grounded model. We propose two possible causes for this drift: small changes in the network coverage which affect our quiet Sun reference velocity due to RV differences between network and the quiet Sun (\citeauthor{Palumbo2017} \citeyear{Palumbo2017}; \citeyear{Palumbo2019}), or RV differences between spots (with penumbrae) and pores (without penumbrae), which are modulated by the changing spot filling factor. In either scenario, more detailed studies of the RV contributions of large and small scale features will be required to elucidate the mechanisms involved.

The different contributions of plage and network to the activity-driven RV variations explains why the calcium H/K activity index does not systematically correlate strongly with RV variations in Sun-like stars on timescales comparable to the magnetic cycle. On highly active stars, where large plage regions dramatically outnumber the small network regions, the plage filling factor will be approximately equivalent to the overall filling factor. We therefore expect the traditional activity indicators, such as $\log{R'_{\rm HK}}$ and optical light curves, to provide a useful proxy for activity-driven RV variations in this regime. On low-activity stars, where the plage and network filling factors are comparable, separating the contributions of plage and network will be necessary to reproduce activity driven RV variations. As the traditional activity indicators are correlated with overall filling factor, they will not provide as useful a proxy of the activity-driven RV variations. For exoplanet RV surveys to be successful for low-activity stars, we must therefore identify correlates for activity region size.

The residuals of our fit still have an RMS spread of over 1.21 m~s$^{-1}$. This additional scatter may be the result of some long-term granulation process \cite{Meunier_et_al_2018}, additional surface velocity flows, additional magnetic effects of network (\citeauthor{Palumbo2017} \citeyear{Palumbo2017}; \citeyear{Palumbo2019}), or unaccounted for systematic variation in the spectrograph on timescales shorter than the solar rotational period. Determining the physical origin of these residual RV variations, identifying correlates for active region size, and verifying that the observed relationships between RV and active region size hold as the Sun enters the active phase of the magnetic cycle will be the subject of future investigations.

\acknowledgments
This work was supported in part by NASA award number NNX16AD42G and the Smithsonian Institution. The solar telescope used in these observations was built and maintained with support from the Smithsonian Astrophysical Observatory, the Harvard Origins of Life Initiative, and the TNG.

This work was performed in part under contract with the California Institute of Technology (Caltech)/Jet Propulsion Laboratory (JPL) funded by NASA through the Sagan Fellowship Program executed by the NASA Exoplanet Science Institute (R.D.H.).

A.C.C. acknowledges support from STFC consolidated grant number ST/M001296/1.

D.W.L. acknowledges partial support from the \emph{Kepler} mission under NASA Cooperative Agreement NNX13AB58A with the Smithsonian Astrophysical Observatory. X.D. is grateful to the Society in Science-Branco Weiss Fellowship for its financial support.

S.S. acknowledges support by NASA Heliophysics LWS grant NNX16AB79G.

L.M. acknowledges the support by INAF/Frontiera through the ``Progetti Premiali'' funding scheme of the Italian Ministry of Education, University, and Research.

 H.M.C. acknowledges the financial support of the National Centre for Competence in Research PlanetS supported by the Swiss National Science Foundation (SNSF)

This publication was made possible through the support of a grant from the John Templeton Foundation. The opinions expressed are those of the authors and do not necessarily reflect the views of the John Templeton Foundation.

This material is based upon work supported by the National Aeronautics and Space Administration under grants No. NNX15AC90G and NNX17AB59G issued through the Exoplanets Research Program. The research leading to these results has received funding from the European Union Seventh Framework Programme (FP7/2007-2013) under grant Agreement No. 313014 (ETAEARTH). 

This work was supported in part by the NSF-REU solar physics program at SAO, grant number AGS-1560313. 

The HARPS-N project has been funded by the Prodex Program of the Swiss Space Office (SSO), the Harvard University Origins of Life Initiative (HUOLI), the Scottish Universities Physics Alliance (SUPA), the University of Geneva, the Smithsonian Astrophysical Observatory (SAO), and the Italian National Astrophysical Institute (INAF), the University of St Andrews, Queen's University Belfast, and the University of Edinburgh.

The HMI data used are courtesy of NASA/SDO and the HMI science team. SDO is part of the Living With a Star Program within NASA's Heliophysics Division.

This research has made use of NASA's Astrophysics Data System.

We thank Director E. Poretti and the entire TNG staff for their continued support of the solar telescope project at HARPS-N.

\facilities{SDO:HMI, TNG:HARPS-N, SORCE:TIM}


\bibliographystyle{yahapj}
\bibliography{references}

\appendix
\section{Calculation of SDO/HMI-derived Quantities}
\label{appA}

In this work, we compute the filling factors of sunspots and plage and the radial velocities (RVs) associated with the suppression of convective blueshift and rotational imbalance. These calculations are based on the methods of \cite{Haywood_et_al_2016}, with some differences. Here, we briefly review the methods of that paper, highlighting the differences in our implementation.

\subsection{Identifying Active Regions}
We identify solar active regions using the same thresholding methods as \cite{Haywood_et_al_2016}. We identify active regions using the line of sight HMI magnetograms. Active pixels have a magnetic field greater than $$\left| B \right| > 3\sigma / \mu $$ where $\sigma = 8 {\rm  G}$, the shot noise per SDO/HMI pixel, and $\mu = \cos{\theta}$, where $\theta$ gives the angular position from the center of the Sun.

To differentiate between dark spots and bright plage, we apply an intensity threshold. We compute the average quiet-sun intensity $I_{\rm quiet}$ by averaging all the inactive pixels identified using the above threshold. Pixels are identified as spots using the intensity threshold of \cite{Yeo_et_al_2013} - that is, if $$I_{ij} < 0.89*I_{\r, quiet}$$

The overall, spot, and plage filling factors are calculated simply by computing the fraction of SDO/HMI pixels corresponding to a given active region type relative to the number of pixels on the solar disk, $N_{sun}$: $$f_{total} = \frac{1}{N_{sun}} \sum_{ij}{W_{ij}}$$

Here $W_{ij} = 1$ if pixel-ij corresponds to an active region, and is 0 otherwise. The same equation may be used to calculate the bright (plage/network) and spot filling factors, $f_{bright}$ and $f_{spot}$: in those cases, $W_{ij} = 1$ if pixel-ij corresponds to falls above or below the intensity threshold described above. We therefore find that $f_{total} = f_{bright} + f_{spot}$.

\subsection{Calculation of Active Region Velocities}

\subsubsection{The convective velocity, $\Delta \hat{v}_{\rm conv}$}

Our calculation of the activity-driven RV shifts differs slightly from that of \cite{Haywood_et_al_2016}. Our calculation of $\Delta \hat{v}_{\rm conv}$ is given by computing the disk-averaged Doppler velocity, $\hat{v}$, and subtracted the disk-averaged quiet-sun velocity, $\hat{v}_{\rm quiet}$: $$\Delta \hat{v}_{\rm conv} = \hat{v} - \hat{v}_{\rm quiet}$$

$\hat{v}$ is the intensity-weighted average of the dopplergram, with the spacecraft velocity and rotation profile ($\hat{v}_{sc}$ and $\hat{v}_{rot}$) removed:

$$\hat{v} = \frac{\sum_{ij}{(v_{ij} - v_{sc,ij} - v_{rot,ij})I_{ij}}}{\sum_{ij}{I_{ij}}}$$

and $\hat{v}_{\rm quiet}$ is the intensity-weighted average over the quiet pixels only:

$$\hat{v}_{\rm quiet} = \frac{\sum_{ij}{(v_{ij} - v_{sc,ij} - v_{rot,ij})I_{ij}\bar{W}_{ij}}}{\sum_{ij}{I_{ij}\bar{W}_{ij}}}$$

where $\bar{W}_{ij} = 1$ for inactive pixels, and is 0 otherwise.

\subsubsection{The photometric velocity, $\Delta \hat{v}_{\rm phot}$}

The photometric velocity is calculated

$$\hat{v}_{\rm phot} = \frac{\sum_{ij}{v_{rot,ij}{(I_{ij}-\hat{K} L_{ij})W_{ij}}}}{\sum_{ij}{I_{ij}}}$$

here $\hat{K}$ is the average quiet-sun intensity at disk center, and $L_{ij}$ gives the limb darkening at the ij-th pixel.

\section{Data}

\noindent Below, we provide the daily-averaged solar telescope and SDO/HMI data products used in our analysis. We include the Julian date of each observation (in days), the solar telescope RV and associated uncertainty, the solar telescope measured $\log{R'_{\rm HK}}$ value, the HMI-derived spot and bright region filling factors, and the HMI-derived photometric and convective velocity shifts. All velocities have units of m~s$^{-1}$. $\Delta \hat{v}_{\rm con, small}$ and $\Delta \hat{v}_{\rm con, large}$ refer to the convective velocity shifts due to active regions with area < 20 ppm and area $\geq$ 20 ppm respectively. Only the first thirty days of observations are listed here: the full table is available online-only as a comma-separated variable (CSV) file.

\begin{center}
\begin{filecontents*}{PublishedData_Short_20ppm.csv}
JD-2457222.5,RV,sigRV,logRHK,fspotE3,fbrightE3,v_phot,v_con,v_con_small,v_con_large
1.405,102.922,0.402,-4.973,0.182,34.169,0.168,5.803,2.732,3.237
10.438,104.131,0.4,-4.967,0.303,42.539,-0.307,7.949,4.203,3.441
10.947,104.602,0.4,-4.965,0.271,43.784,-0.214,8.386,4.596,3.577
11.968,104.754,0.401,-4.963,0.386,45.955,-0.002,9.135,5.402,3.73
12.965,105.195,0.401,-4.965,0.356,46.37,0.093,9.407,5.679,3.819
14.035,105.606,0.397,-4.963,0.264,44.349,0.175,9.084,5.412,3.847
14.975,105.808,0.393,-4.962,0.246,44.673,0.289,8.23,4.611,3.911
15.571,105.224,0.39,-4.961,0.299,45.293,0.299,7.575,4.027,3.842
17.042,103.588,0.399,-4.958,0.595,44.351,0.424,6.473,3.242,3.657
17.512,104.507,0.39,-4.957,0.756,43.556,0.492,6.238,3.111,3.631
19.468,103.439,0.4,-4.97,1.277,41.337,0.246,6.611,3.348,3.463
19.716,103.975,0.402,-4.971,1.316,41.01,0.133,6.68,3.337,3.448
22.471,100.788,0.4,-4.978,1.136,39.226,-0.554,6.757,2.66,3.507
22.504,101.162,0.4,-4.978,1.129,39.216,-0.553,6.746,2.649,3.51
29.425,103.184,0.516,-4.986,0.046,29.373,0.111,4.574,1.476,3.214
30.006,102.178,0.45,-4.98,0.049,29.155,0.083,4.378,1.314,3.145
31.351,101.838,0.404,-4.972,0.227,30.244,0.017,4.137,1.107,3.066
32.267,99.889,0.404,-4.979,0.461,31.03,0.15,4.124,1.186,3.064
33.261,102.153,0.399,-4.976,0.697,32.619,0.313,4.705,1.956,3.076
33.751,103.12,0.393,-4.973,0.845,33.721,0.341,5.182,2.501,3.031
34.579,104.932,0.4,-4.97,1.064,34.662,0.238,6.219,3.361,3.101
37.195,105.65,0.403,-4.961,1.66,39.151,-0.798,8.446,4.431,3.256
38.04,104.555,0.402,-4.967,1.595,40.119,-0.899,7.969,3.904,3.162
38.994,102.762,0.406,-4.971,1.465,42.171,-0.722,7.135,3.09,3.324
40.011,101.493,0.408,-4.971,1.231,44.358,-0.363,6.17,2.417,3.4
40.953,101.741,0.403,-4.97,0.436,41.352,0.017,5.607,2.281,3.342
41.938,102.048,0.406,-4.968,0.051,37.677,0.039,5.978,2.329,3.688
42.896,102.305,0.408,-4.976,0.05,38.559,0.037,5.814,2.229,3.625
43.918,102.659,0.403,-4.976,0.069,38.382,0.09,5.732,2.249,3.574
44.981,103.054,0.406,-4.973,0.103,36.822,0.106,5.632,2.247,3.49
\end{filecontents*}
\csvautolongtable[table head= JD - 2457222.5 & RV & $\sigma_{RV}$ & $\log{R'_{\rm HK}}$ & $f_{spot} \times 10^3$ & $f_{bright} \times 10^3$ & $\Delta \hat{v}_{\rm phot}$ & $\Delta \hat{v}_{\rm conv}$& $\Delta \hat{v}_{\rm conv, small}$ & $\Delta \hat{v}_{\rm conv, large}$ \\ \hline, table foot =\\ \vdots & \vdots & \vdots & \vdots & \vdots & \vdots & \vdots & \vdots & \vdots & \vdots]{PublishedData_Short_20ppm.csv}
\end{center}

\end{document}